\documentclass[a4,notitlepage,twocolumn]{revtex4-1}
\usepackage{graphicx}
\usepackage{epstopdf}
\usepackage{longtable}

\begin{document}

\title{Solving the dark matter problem by dynamic interactions}
\author{Werner A. Hofer}
\affiliation{Newcastle University, Newcastle upon Tyne NE1 7RU, UK}
\email{werner.hofer@ncl.ac.uk}

\begin{abstract}
Due to the renewed interest in dark matter after the upgrade of the large hadron collider and its dedication to dark matter research it is timely to reassess the whole problem. Dark matter is one way to reconcile the discrepancy between the velocity of matter in the outer regions of galaxies and the observed galactic mass. So far, no credible candidate for dark matter has been identified. Here, we develop a model accounting for observations by rotations and interactions between rotating objects analogous to magnetic fields and interactions with moving charges. The magnitude of these fields is described by a fundamental constant of the order 10$^{-41}$kg$^{-1}$. The same interactions can be observed in the solar system where they lead to small changes of planetary orbits.
\end{abstract}


\maketitle
Observations in radio astronomy of the 21cm hydrogen line in interstellar gas and its broadening due to relative motion of hydrogen clouds in interstellar space within galaxies have been employed since the 1960s to measure their rotational velocities relative to a galactic centre\citep{1,2,3}. From the distribution of stellar mass in the galaxies it was expected that the rotational velocity would decrease roughly with the inverse square root of the distance from the centre\citep{4}. However, observations by V. Rubin established that the velocity does not decrease, but either remains constant with increasing distance, or even slightly increases\citep{2,4}, see Figure 1. This fact poses a substantial challenge for our current understanding of nature at this scale, as it either means that we do not understand, what a large part of our universe is composed of (the mass unaccounted for is more than five times the baryonic mass in the universe\citep{5,5.1}), or that we do not understand how mass interacts (neither Newton’s nor Einstein’s formulation of the law of gravitation accounts for the discrepancy). This discrepancy was originally termed the "missing matter" problem, which has since morphed into the "dark matter" problem \citep{5.1}, as it is thought that it might find its solution by a genuinely new form of matter in the universe which does not interact in conventional ways. The concept of dark matter, as Milgrom pointed out\citep{6}, is based on three separate assumptions: (i) the force governing the dynamics of interstellar hydrogen is gravity, (ii) the gravitational force depends on the source of the gravity field and the mass of the particle, and (iii) Newton’s second law. Milgrom modified Newton’s second law to read:
\begin{equation}
m_g \mu\left(\frac{a}{a_0}\right) \mathbf{a} = \mathbf{F}
\end{equation}
where $m_g$ is the mass of hydrogen, $\mu$ is a function depending on acceleration $a$ and a constant $a_0$,
and $\mathbf{F}$ is the force due to acceleration $\mathbf{a}$. For accelerations much larger than
$a_0 = 2 \times 10^{-10}$ms$^{-2}$ Newtonian dynamics is restored, while for smaller acceleration the relation leads to a constant orbital velocity $V = \sqrt[4]{G M a_0}$, where $G$ is the gravitational constant and $M$ the mass of a galaxy\citep{6}.
\begin{figure}
\begin{center}
\includegraphics[width=\columnwidth]{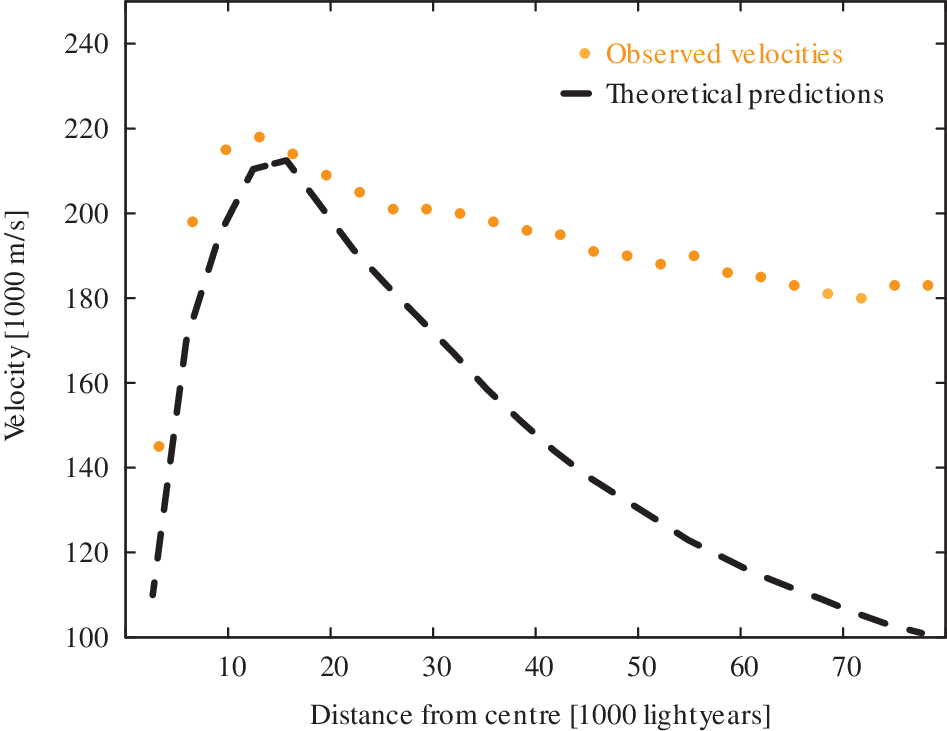}
\caption{Observations and theoretical predictions for the galaxy NGC 2903. Observed (orange dots) and predicted (broken black line) velocity of interstellar hydrogen in the galaxy. The predictions are based on gravity and the observed stellar mass (data from Ref. \citep{4}).}\label{F1}
\end{center}
\end{figure}
However, Eq. (1) is only a "working formula, of limited validity [ . . . ] and does not constitute a theory"\citep{7}. Moreover, if one considers the whole of physical theory and not just the - essentially static - laws of gravitation formulated today, then assumption (i) may not be strictly valid. The reason for such an assumption can be found in electrodynamics\citep{8}, where charges interact through two separate fields: they interact either through electrostatic Coulomb fields, equivalent to gravitational interactions of mass, or through magnetic fields, which currently have no equivalent in the theory of mass dynamics. It will be shown that such a field, which we call the rotor field $\mathbf{R}$, leads to identical results, suitable to account for the dark matter problem.

The concept of rotor fields and forces is similar to modified Newtonian dynamics insofar it provides a solution to the problem in a change of interactions of moving mass rather than in additional mass components. Given that no credible candidates for these additional components have been found or can even be envisaged at present, an approach based on interactions is at once safer and more grounded in scientific reality. Compared to modified Newtonian dynamics the concept of rotor fields and rotor forces leaves the general validity of Newtonian mechanics unquestioned, i.e. it does not try to assert that basic physical laws are only valid for a limited range of accelerations. Moreover, by providing a general framework for interactions of rotating mass via fields and forces due to rotations, it also allows for a generalization of the concept to other environments, for example the solar system. So it has the additional advantage of being more generally applicable.

Galaxies come in many different shapes. Here, we concentrate on symmetrical disk-shaped galaxies. The reason for this restriction is that excellent data of such galaxies exist in the work of Begeman et al.\citep{4}, and that simplifications are necessary to keep the new theoretical model as transparent and simple as possible. A galactic disk is a two dimensional system, see Figure 2. The following model of stellar dynamics is based on two discrete assumptions: (i) Mass in angular rotation around the galactic centre generates a rotor field which is radiated outward in the plane of the galaxy, and (ii) the field amplitude decreases linearly with increasing distance from the centre. The second assumption is Gauss’s law applied to a two dimensional system\citep{8}. Since the circumference of a circle at radius $r$ is $2 \pi r$, the field amplitude $\mathbf{R}$ must be inverse proportional to $r$. Such systems are not unknown in electrodynamics, as the field of antennas will commonly be very directional\citep{8}. The field $\mathbf{R}$ for positive rotation can then be described by:
\begin{equation}\label{2e}
\mathbf{R}\left(r\right)= \mu_R \frac{\mathbf{e}_z}{r}\int dr' \rho(r')2 \pi r' V(r')
\end{equation}
Here, $\mu_R$ is the rotor constant, which will be determined from astronomical data, $\mathbf{e}_z$
is a unit vector in $z$-direction, $\rho(r')$
 is the two-dimensional mass density, and $V(r')$ the angular velocity of the mass distribution. As the mass density of a galaxy is usually concentrated near the centre and decreases with the square of the distance from the centre\cite{4}, we may further simplify the equation by assuming that most of the galactic mass $M$ is contained within a certain radius $r_0$, and that the velocity distribution of this mass can be replaced by a constant $V_0$. This simplifies Eq. (2) to:
 \begin{equation}\label{3e}
 \mathbf{R} = \mu_R \frac{\mathbf{e}_z}{r} M V_0
 \end{equation}

In electrodynamics the interaction between a magnetic field and charge in motion is described by the Lorentz force\citep{8}. We assume that a similar law holds for the interaction of a rotor field and mass in motion. For a velocity vector directed in $\varphi$-direction and due to circular rotation around the galactic centre $\mathbf{V}(r) = \omega(r) \mathbf{e}_{\varphi}$ the resulting rotor force $\mathbf{F}_R$ on unit mass will be directed inward along radial vector $\mathbf{e}_r$, and equal to the centrifugal forces directed outward, if $\omega(r) = V_0/r$
\begin{equation}\label{4e}
\mathbf{F}_R = \mathbf{R}(r) \times \mathbf{V}(r) = - \mu_R \frac{M V_0^2}{r} \mathbf{e}_r
\end{equation}
for a suitably chosen value of the rotor constant $\mu_R$.
The result is the same as in modified Newtonian dynamics: while gravity as described by Newton’s or Einstein’s theory of gravitation will lead to a decreasing angular velocity as the distance from the centre increases, this model of rotor fields and dynamic interactions yields a flat velocity distribution. The rotor constant $\mu_R$ can be calculated from existing data with the relationship between velocities $V$, galactic mass $M$, and acceleration constant $a_0$.  For many galaxies the observed velocity is between 100 and 300 km/s. In our simple model this implies that $\mu_R$  is between $10^{-40}$kg$^{-1}$ and $10^{-42}$kg$^{-1}$. The forces related to these interactions are therefore comparatively small.
\begin{figure}
\begin{center}
\includegraphics[width=\columnwidth]{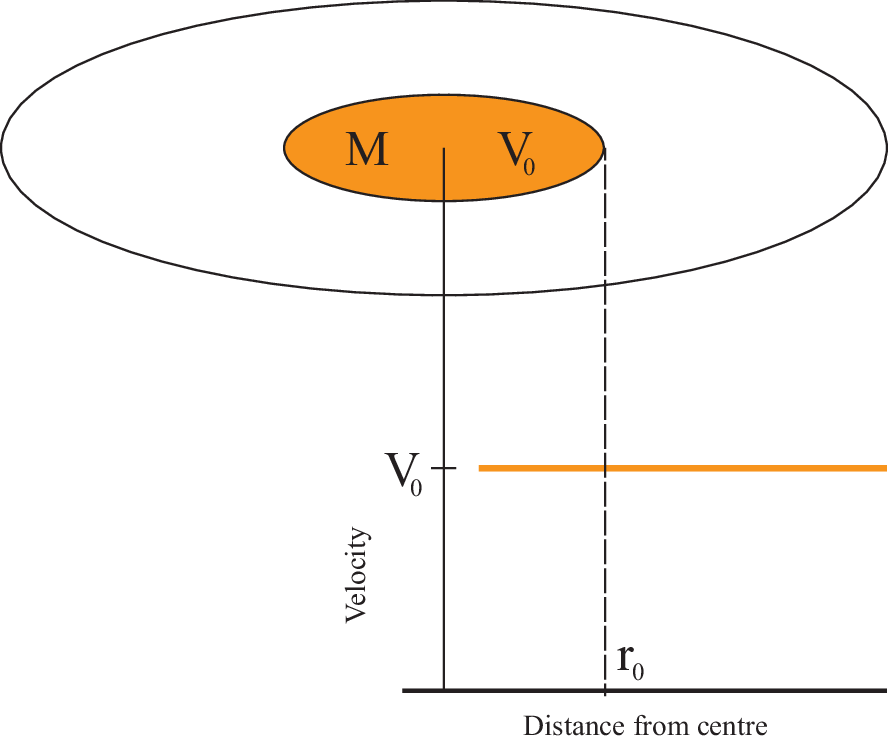}
\caption{ Effect of rotor fields in a simple model. The spiral galaxy is simplified to a disk. The orange circle indicates the cutoff radius for the central rotating mass, the velocity distribution due to rotor fields and their interaction will be flat in the outer regions of the disk.}\label{F2}
\end{center}
\end{figure}
While the model explains, why rotational velocity remains flat in the outer regions of a disk shaped galaxy, it does not account for the Tully-Fisher relation between the mass of a galaxy and rotational velocity\cite{9}. Here, it has to be considered that mass rotating near the centre of a galaxy generates a rotor field which is radiated outward. Then the question arises, whether rotor fields will influence rotations also near the centre. There has to be some radius $r' = r_0/x$, where the rotor fields of matter rotating inside will determine the dynamics of mass rotating outside, while inside this radius dynamics will be determined mainly by gravitational attraction. The mass inside $r'$ is $M/x^2$, therefore the equivalence of centrifugal and gravitational forces leads to:
\begin{equation}\label{5e}
V_0^2 = \frac{G M}{r_0 x}
\end{equation}
Empirically, it is known that $V_0$ increases with increasing $r_0$ of a galaxy. In the simplest case the relation between $V_0$ and $r_0$ will be linear. It is also plausible that the radius $r' = r_0/x$ will not vary too much for different galaxies, that is, it will not increase in the same way as $r_0$. To first approximation this means that $x$ is proportional to $r_0$ . Since $r_0$ is proportional to $V_0$, the product $x r_0$ is proportional to $V_0^2$. Setting the constant equal to $a_0^{-1}$, we recover the previous result. However, it is likely that $a_0$ will vary for different galaxies, and astronomical data show that the velocity does not exactly follow a power four law\citep{9}, i.e. this derivation of the Tully-Fisher relation is only an approximation.

If rotor fields and their interactions are real, then they must apply to all rotating masses, also to mass within the solar system. Here, the dominant mass is the mass of the sun. At $1.99 \times 10^{30}$kg the solar mass\cite{10}  is about ten orders of magnitude smaller than the mass of a galaxy; effects within this system on planetary orbits will consequently be much smaller. The mass of the sun rotates with a period varying between 25 days (equator) and 35 days (near the poles) \citep{10}. For the following we assume rigid rotation with a period of 26 days, and we also assume that the density of solar mass is constant. For $\omega_S = 4.45 \times 10^{-7}$s$^{-1}$ and a solar radius $r_S = 6.96 \times 10^8$m the rotor field in the plane perpendicular to the solar axis of rotation will be:
\begin{equation}\label{6e}
\mathbf{R}(r) = \mu_R \frac{\mathbf{e}_z}{r} 3.63 \times 10^{32} kg s^{-1}
\end{equation}
The force on a unit mass at the position of the mercurial or terrestrial orbit is consequently $\mu_R \cdot 4.78 \times 10^{25}$N/kg and $\mu_R \cdot 1.15 \times 10^{25}$N/kg, respectively. To determine the effect on a planetary orbit we assume that the additional force will manifest itself as an additional centripetal force which will lead to a slight increase in the orbital velocity of mass. For $\mu_R = 10^{-40}$kg$^{-1}$ we obtain an additional velocity of 16.6 mm/s (Mercury) and 13.1 mm/s (Earth) \citep{11}. From the given velocities we calculate that the orbital shifts in one terrestrial year are 1.87 and 0.57 arc seconds, respectively.
\begin{figure}
\begin{center}
\includegraphics[width=\columnwidth]{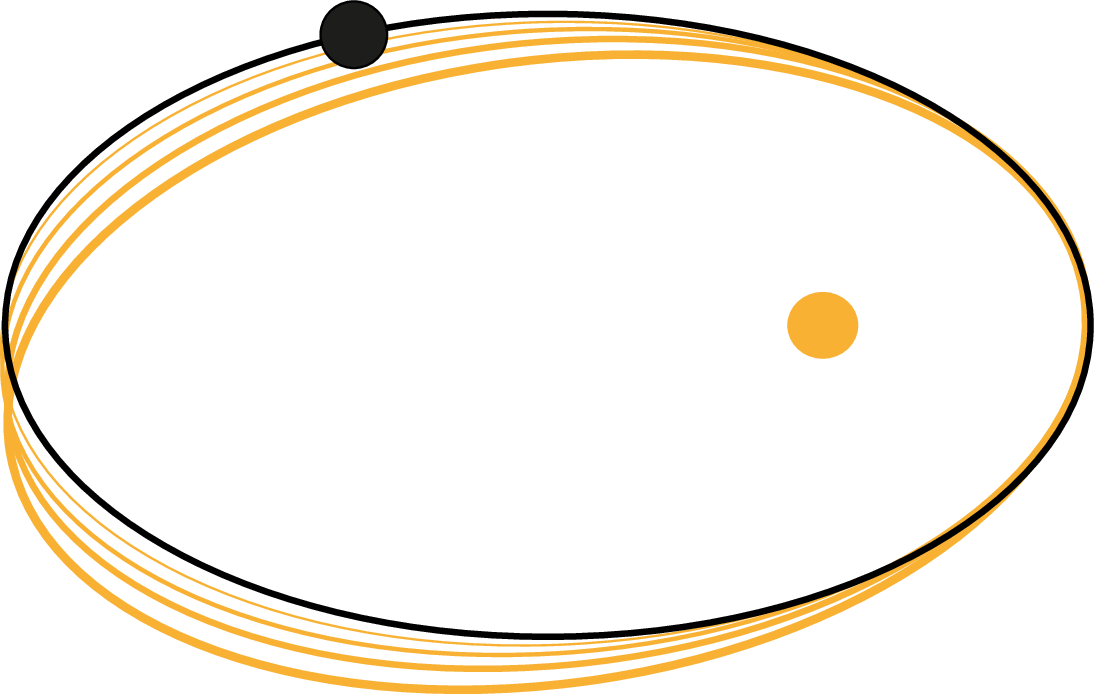}
\caption{Effect of solar rotation on planets of the solar system. Black: stable elliptical orbit as a consequence of gravitational pull from a central mass. Orange: Mercury’s elliptical orbit will rotate around the sun, the rotation is visible as a gradual advance of the perihelion, the point of closest approach to the sun. Note that most of the advance is due to the gravitational pull of solar planets, only a small fraction of  43 arc seconds remains unexplained in Newton’s gravitational model.}\label{F3}
\end{center}
\end{figure}
The value for the rotation of the mercurial orbit at 187 arc seconds per century is about four times larger than the value calculated in General Relativity \citep{12}. Assuming that the perihelion advance of Mercury, shown in Figure 3, is due to rotor fields would allow us to calibrate the rotor field constant. For 43 arc seconds per century the value of $\mu_R = 0.53 \times 10^{-41}$kg$^{-1}$. This value is well within the range determined from the dynamics of galaxies. The advance of the terrestrial orbit will then be 13 arc seconds or about three times larger than the effect calculated with general relativity.

An identical calculation can be made for the Venutian orbit \citep{14}, which yields a result of 12 arc second for the rotation of the orbit. It has indeed been obsserved that the nodes of Venus move by about 10 arc seconds per century \citep{15}, an observation which cannot be explained by General Relativity and is indeed accounted for - within the errors likely to come from the simplification of this model - in the model of rotor fields and rotor forces. Since the effect of rotor fields, as explained above, will lead to flat velocity distributions with increasing distance from the centre of the solar system, we expect similar deviations from current theories on gravitation also for the outer regions of the system. To summarize, the model of rotor fields and rotor forces can account for three separate effects:
\begin{enumerate}
\item The flat velocity distributions of interstellar hydrogen in galaxies.
\item The advance of the perihelion of Mercury.
\item The advance of the nodes of Venus.
\end{enumerate}
General Relativity, by contrast, can only account for one of the three effects (Mercury's perihelion) and, most importantly, fails completely to explain the rotational velocities within galaxies.

It is surprising that using the missing matter problem as a starting point one can account for the problem by postulating a new form of mass interactions, which has observable consequences in our immediate vicinity. The predicted changes of planetary orbits also agree reasonably well with astronomical data and with theoretical predictions made in a completely different context. At this point the question is legitimate, whether the model of rotor fields and forces aims at providing an alternative explanation for the advance of planetary orbits to the one currently accepted in the scientific community, i.e. the explanation forwarded by the concept of General Relativity. Strictly speaking, this question cannot be answered at present. We show only with this model that the question of the observed rotational velocities in distant galaxies may not be unrelated to the question how planetary orbits are affected by the rotation of the sun. Whether it is the rotation of solar mass or the distortion of the flat space time geometry due to solar mass which causes the advance of mercury's orbit, will have to be addressed in the future. However, it is quite astonishing that the fundamental constant governing one effect - rotational velocities in galaxies - is within a range which makes it plausible that it may also govern the other effect - rotation of planetary orbits.

If it is established that there is indeed a relation between these two effects then it seems possible that this model will open up astrophysics to a truly dynamical analysis of mass interactions in the same way classical electrodynamics was made possible by the discovery of magnetic fields and their interactions. However, it cannot be expected at this point that the application of this model will be able to account for every astronomical observation of stellar dynamics without additional development. Moreover, as the estimate of the effect of solar rotation showed, rotor fields arise due to orbital velocity of mass around the galactic centre and rotation of mass in individual stellar objects. In practice, this will lead to a large variability in the dynamics of individual galaxies.

In summary we have shown that a new mode of interactions of mass in motion leads to the prediction of flat velocity distributions in galaxies and a dependency of this velocity on galactic mass. The results agree with astronomical observations and account for the problem known as the dark matter problem. Similar interactions play a role in the solar system, where they lead to predictions for the advance of planetary orbits equal to predictions in General Relativity.

\section*{Acknowledgements}
The author acknowledges EPSRC support for the UKCP consortium, grant No. EP/K013610/1.


\begin{references}
\bibitem{11}
For the calculations we assumed circular orbits, with $r_M = 5.79 \times 10^{10}$m and $\omega_M = 1.32 \times 10^{-7}$s$^{-1}$ for Mercury, and $r_E = 1.50 \times 10^{11}$m and $\omega_E = 3.17 \times 10^{-8}$s$^{-1}$ for Earth.
\bibitem{14}
For the calculations of Venus we assumed a circular orbit with $r_V = 1.08 \times 10^{11}$m and $\omega_V = 5.15 \times 10^{-8}$s$^{-1}$.
\end{references}
\end{document}